\documentclass[linenumbers,preprint,12pt,authoryear]{elsarticle}

\usepackage{amsmath,amssymb,amstext}
\usepackage{subfigure}

\usepackage{float}

\usepackage{multirow}
\usepackage[utf8]{inputenc}
\usepackage{lmodern}

\usepackage{mathtools,nccmath}%

\usepackage{environ}

\usepackage{natbib}

\usepackage{hyperref}

\usepackage{comment}

\usepackage{xcolor}

\newcommand{\lc}{\ell_{\rm c}}

\newcommand{\rl}{\mathcal{R}}

\journal{Astroparticle Physics}

\begin{document}

\begin{frontmatter}

\title{ Revisiting Propagation Delays of Ultra-High-Energy Cosmic Rays from Long-lived Sources}

 \author[label1,label2,label3]{Rostom Mbarek}
 \affiliation[label1]{organization={Joint Space-Science Institute, University of Maryland},
             city={College Park},
           postcode={20742},
            state={MD},
             country={USA}}

 \affiliation[label2]{organization={Department of Astronomy, University of Maryland},
             city={College Park},
           postcode={20742},
            state={MD},
             country={USA}}
 \affiliation[label3]{organization={Astrophysics Science Division, NASA Goddard Space Flight Center},
             city={Greenbelt},
           postcode={20771},
            state={MD},
             country={USA}}

\author[label4,label5]{Damiano Caprioli}

\affiliation[label4]{organization={University of Chicago, Department of Astronomy \& Astrophysics},
            address={5640 S Ellis Avenue},
             city={Chicago},
           postcode={60637},
            state={IL},
             country={USA}}
\affiliation[label5]{organization={Enrico Fermi Institute, The University of Chicago, Chicago},
             city={Chicago},
           postcode={60637},
            state={IL},
             country={USA}}

\begin{abstract}
We revisit the time delay incurred during ultra-high energy cosmic ray (UHECR) propagation over cosmological distances and its potential impact on the correlation between UHECR directions of arrival and long-lived sources (i.e., with duty cycles of order of Myr, such as Active Galactic Nuclei (AGNs) and starburst galaxies), the UHECR chemical composition, and extragalactic magnetic field constraints.
We propagate particles in different magnetic field configurations, spanning over an extended range of particle Larmor radii and magnetic field coherence lengths, also including attenuation losses.
We conclude that UHECR delays could easily be comparable to (and longer than) AGN duty cycles, effectively erasing the correlation between known AGNs and UHECR anisotropies.
We finally consider how strong constraints on the chemical composition of the heaviest UHECRs could enable a better characterization of extragalactic magnetic fields.
\end{abstract}


\end{frontmatter}


\section{Introduction}
Pinpointing the sources of the Ultra-High-Energy Cosmic Rays (UHECRs) involves finding a correlation between their time/directions of arrival and the distribution of potential sources. 
The correlation with transient or short-lived sources, such as gamma-ray bursts and newly-born millisecond neutron stars, is washed away by the delay induced by extragalactic propagation \citep[e.g.,][]{miralda-escude+96,murase+09,kalli+11,takami+12,harari+21,vanVliet+21}, and by the delay accumulated in our Galaxy, which is likely $\gtrsim$1000~yr for particles with rigidities of $10^{19}$V or less
\citep[e.g.,][]{alvarez-muniz+02, murase+09, takami+12}.

For Active Galactic Nuclei (AGNs), the same arguments apply to UHECRs produced during flaring phases \citep[e.g.,][]{biermann+87,dermer+09,murase+12,farrar+09}, but comparing propagation delays with their much longer duty cycles (of the order or Myr or more, as outlined below) requires modeling extra-galactic propagation.

In this work, we consider a large range of possible properties of the magnetic fields in/around UHECR sources, in galaxy clusters, and on cosmological scales to examine potential UHECR delays during propagation, and their impact on the detected UHECR chemical composition. 
We find that UHECR propagation generally introduces delays comparable to AGN duty cycles, rendering statistical correlations and anisotropy studies involving these sources quite challenging;
moreover, these conclusions likely extend to starburst galaxies, since the periods of enhanced star formation could be comparable to AGN duty cycles, for instance when fed by major galaxy mergers. 
Star formation can be quenched either by AGN or stellar feedback; 
in the former case there is a direct physical connection between the two duty cycles, in the latter quenching should kick in when massive stars go supernovae, which is still in the Myr ballpark. 
Within the order of magnitude uncertainty that applies to AGN duty cycles, we conclude that the arguments in this paper should apply to scenarios in which UHECR sources are either starburst galaxies or AGNs.

\subsection{AGN duty Cycles}
The origin of an AGN duty cycle $t_{\rm AGN}$, i.e., the time interval for which a supermassive black hole is active, is quite uncertain and likely depends on the mass and luminosity of the host galaxy. 
Several studies have focused either on local radio galaxies with $z<0.1$ \citep[e.g.,][]{shabala+08} or on radio-loud FR-I and FR-II galaxies \citep[e.g.,][]{hardcastle+19,shabala+20} and found observational evidence that place $t_{\rm AGN} \gtrsim 10^{6}$-$10{^7}$yr. 
We can also get clues on $t_{\rm AGN}$ by constraining the average lifetime of quasars, using three main methods.
The first is the \textit{transverse proximity effect}, where changes in radiation, and thus photoionization rates, affect the Ly$\alpha$ forest of a bright quasar.
Such an impact on the photoionization rates informs us on quasar duty cycles \citep[e.g.,][]{kirkman+08,schmidt+17,bosman+20}, and gives $t_{\rm AGN} \gtrsim 10^{5}$-$10^{7}$yr. 
The second method uses the \textit{extent of ionized nebulae around quasars} to probe their reprocessed radiation field, and hence their radiative history, with $t_{\rm AGN} \gtrsim 10^{6}$-$10^{7}$yr \citep[e.g.,][]{trainor+13,borisova+16}.
The third method, dubbed \textit{quasar clustering}, compares the relative abundance of quasars with their host halos \citep{martini+01,haiman+01}; with somehow large uncertainties, it returns $t_{\rm AGN} \gtrsim 10^{6}$-$10^{8}$yr \citep{martini+01,hopkins+05,shen+09}. 
AGN duty cycle measurements are hence not fully constrained, but they should lie in the range $t_{\rm AGN}\gtrsim 10^{5}-10{^7}$yr and not exceed $\sim 10^8$yr.

\subsection{Magnetic Fields and UHECR Propagation}\label{sec:introB}

Constraining the properties of magnetic fields between UHECR sources and Earth is of paramount importance for their propagation and for evaluating the effects of propagation delays. 
In this regard, the effects of the galactic magnetic field (GMF), intergalactic magnetic field (IGMF), the intra-cluster medium (ICM) $B$-fields, and even $B$-fields in voids should be important.

UHECRs can experience important deflections because of the GMF \citep[e.g.,][]{takami+12b,jansson+12,CRpropa16,unger+17,unger+24}, especially because GMFs can easily reach $B\sim 1$-$10 \mu $G.
In this regard, the effects of the GMF toroidal component \citep{sun+08,pshirkov+11}, and poloidal and random components \citep{jansson+12,farrar+19} have been found to potentially cause significant deflections.
Generally speaking, the structure of the GMF remains poorly constrained, but its effects on UHECR deflection should be inevitable, with delays of the order of $\sim 10^3$yr for our Galaxy \citep[e.g.,][]{murase09}.

As for the extragalactic components, numerous studies have attempted to constrain the properties of intergalactic magnetic fields (IGMFs) \citep[e.g.,][]{neronov+10,tavecchio+10-igmf,tavecchio+11,huan+11,taylor+11,finke+15} and their corresponding coherence lengths \citep[e.g.,][]{vovk+12}.
However, the scarcity of observational data and the difficulties associated with measurements of $B$-fields \citep{batista+21} have made this challenging. 
Moreover, in the context of quasars, outflows such as jets can inject quite substantial fields \citep[e.g.,][]{kulsrud+08,vallee+11,ryu+12} in the IGM \citep{furlanetto+01}. 

As for the ICM, it may also be populated with non-negligible fields, especially that filaments of galaxy clusters are found to have $B$-fields reaching $B\sim0.1$-$10$nG \citep[e.g.,][]{vazza+17b}, with maximum values based on magnetic draping reaching a few $\mu $G \citep{pfrommer+10}. 
Faraday rotation observations also suggest that $B$-fields can reach $\sim \mu$G levels \citep[e.g.,][]{clarke+01,carilli+02,govoni+04}.
The local galaxy supercluster, on the other hand, has been found to have a statistically significant ``Faraday screen'' acting on radio waves, which would return a field as large as $B\sim 0.3 \mu$G in the local Universe \citep{vallee+02}. 
Similarly, the plane of the Coma supercluster can reach 0.5-1.5 $\mu$G, declining to about 0.03-0.5 $\mu$G on $\gtrsim 1$ Mpc levels \citep{brunetti+01}.
Other work demonstrated that radio halos in the ICM are diffuse Mpc-scale synchrotron sources produced by GeV electrons interacting with $\mu$G-scale magnetic fields \citep{cassano+10,cuciti+22}.
Finally, simulated $B$-fields in the local supercluster find that they can reach $\sim 10$nG \citep{ryu+08}. 

It is worth mentioning that regions with indications of UHECR excesses have been reported to intersect the supergalactic plane \citep[e.g.,][]{AUGER18,AbdulHalim:2023w+,Kim:2023vZ,AUGER24c,globus+19}, and that the UHECR dipole anisotropy could be associated with the orientation of the Galaxy relative to the supergalactic plane \citep{globus+17b,globus+19,ding+21,bister+24a}.
Other studies suggested that cluster $B$-fields were $\sim 3 \mu$G based on synchrotron cooling of populations of secondary electrons inferred from the correlation between the radio and thermal X-ray luminosity of radio emitting galaxy clusters \citep{kushnir+09}.
Within the context of UHECR propagation over cosmological baselines, it has been proposed that $B$-fields in voids are the most decisive component \citep{batista+21}, and an upper limit for their values could be set at $\lesssim 1$nG \citep[e.g.,][]{blasi+99,kronberg+07,planck+16,pshirkov+16}.

Despite the substantial body of work that has attempted to constrain the properties of $B$-fields in galaxies, galaxy clusters and beyond, their basic properties in regions that separate us from UHECR sources remain poorly known. 
Previous studies have investigated the role of propagation delays in shaping the observed UHECR flux and features. 
\citet{matthews+18} examined whether past activity in nearby AGNs, Centaurus A and Fornax A, could explain the present-day UHECR flux, accounting for a possible efficient past AGN activity and the time delays accumulated during propagation. 
\citet{eichmann+22b,eichmann+23} analyzed how source lifetimes and magnetic deflections affect spectral hardening and composition evolution, highlighting the importance of extended propagation times in shaping observed spectra.
In the context of the Local Supercluster, \citet{mollerach+19, mollerach+22} explored the effects of CR diffusion including finite source ages, emphasizing their influence on spectral features, time delays, and anisotropy. 
Adding to this picture, \citet{fargion+18} estimated that AGN flare episodes are unlikely to correlate with UHECR arrival times, given potential propagation delays reaching up to $10^6$  years. 
In general, even delays as small as $\lesssim 1$Myr make associations of transient activity and UHECR detections very challenging. 

Collectively, these works have deepened our understanding of how magnetic fields, source characteristics, and propagation effects modulate UHECR observables.
Numerical codes, such as {\tt SimProp} \citep[e.g.,][]{aloisio+12} and {\tt CRPropa} \citep[e.g.,][]{batista+15,CRPROPA22}, have been developed to study UHECR propagation.
In this work we develop our own agile framework for calculating propagation delays, which incorporates the effects of the most updated theory of CR scattering and can be applied to several different environments: inside/around sources, within host galaxies and clusters, and across cosmological distances. 
Specifically, we attempt to bracket current uncertainties by propagating particles without assuming diffusion a priori, exploring an unprecedentedly-wide range of magnetic field strengths, spectral indices, and coherence lengths. 
We also account for energy losses due to photomeson production and photodisintegration from interactions with the cosmic microwave, infrared, and optical backgrounds \citep{dole+06}. 

While earlier studies have examined UHECR time delays in more restricted scenarios with specific assumptions and/or considering a limited region of the parameter space \citep[e.g.,][]{bhattacharjee+00,das+08,davoudifar+11}, our  overarching goal is to assess whether UHECR propagation is generally subject to delays that exceed also long-lived sources, such as AGN and starburst galaxies, with their  multi-Myr duty cycles.
In addition, we show how a precise measurement of the UHECR chemical composition can place tighter constraints on the properties of extragalactic magnetic fields.

\section{Particle propagation}\label{sec:propagation}

We consider a simple setup where test particles are propagated with a Boris pusher \citep[e.g.,][]{birdsall+91} with different initial pitch angles $\mu$ in a prescribed magnetic field $\mathbf{B}$. 
A rigidity-dependent time step is chosen to ensure proper resolution of particle gyrations, with more than five steps per gyroperiod. While the magnitude of the magnetic field, $B$, remains constant during particle propagation, a turbulent component is added to alter its direction. 
The field direction changes by an angle $\theta$, following a random walk every coherence length $\lc$, which characterizes the local magnetic turbulence. 
The angle $\theta$ is drawn from a Gaussian distribution centered at zero with a standard deviation of $\pi/2$, simulating sharp, strong magnetic field bends ($\delta B/B \sim 1$).
The propagated particles represent different species, since propagation is dependent on rigidity only (except for losses). 

Particle transport has been extensively studied in turbulent boxes with magnetohydrodynamical (MHD) simulations \citep[e.g.,][]{beresnyak+11,cohet+16, dundovic+20}, but this approach is unfeasible for the vast range of scales involved in UHECR propagation. 
Instead, frameworks where particle diffusion is simulated on synthetic fields are routinely adopted to study UHECR transport \citep[e.g., {\tt CRPropa},][]{CRpropa16}. 
Our approach, which does not assume diffusive transport but only the random walk of the magnetic field lines,  allows us to explore a wide range of Larmor radii $\rl$ much smaller/larger than the scattering mean free path (not possible in a diffusive approximation), $B$ coherence length $\lc$, and distance to the source.
Though it ignores the actual cosmological distribution of matter in the Universe, our quite simple framework is sufficient to investigate the effects of typical strengths and coherence lengths on the propagation delays, obtained by comparing the UHECR travel time $t$ with the ballistic time $t_b$ from the source, i.e., the photon (or neutrino) travel time.
We now consider different transport regimes, depending on whether $\mathcal{R}\gtrless l_c$.

\begin{figure}[htbp]
    \hspace{-2.1cm} 
    \includegraphics[width=1.3\textwidth]{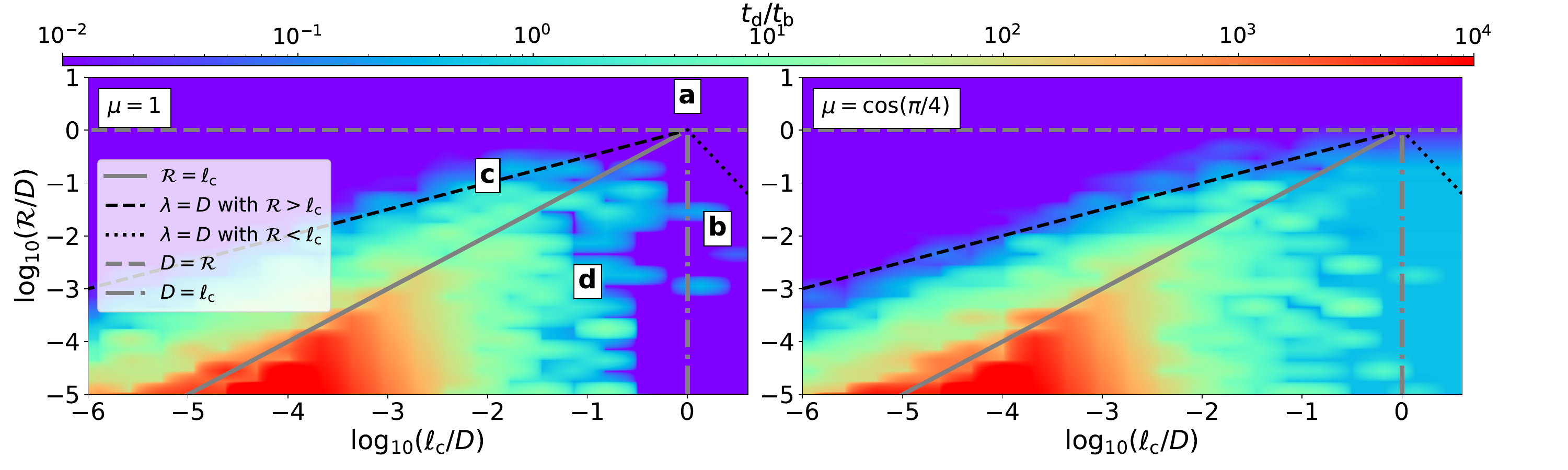}
    \caption{Time delay maps of particle propagation as a function of the particle Larmor radius $\mathcal R$ and coherence length $\lc$ normalized to the source distance $D$. The color bar denotes the ratio of the time delay $t_d = t-t_b$ and the photon propagation time $t_b$. The solid grey line characterizes the boundary $\mathcal R = \lc$. Particle Larmor radii are also plotted for the limiting case $\lambda = D$ (dashed black for $\mathcal R > \lc$ and dotted line for $\mathcal R < \lc$). The different maps show the impact of the $B$-field setup and initial pitch angle. Particle losses are not accounted for in these maps.}
    \label{fig:gyro}
\end{figure}

\subsection{Regime with $\rl > \lc$}\label{sec:rglc}

This regime is particularly interesting for the transport of high-energy particles, as in small B fields Larmor radii could easily exceed coherence length scales.
For $\rl \geq \lc$, particles are scattered on scales smaller than their Larmor radii resulting in a scattering mean free path $\lambda$,
\begin{equation}\label{eq:lambda}
    \lambda \sim \lc^{1-\delta}\rl^{\delta}
\end{equation}
with $\delta = 2$ \citep[e.g.,][]{shalchi+09,plotnikov+11,plotnikov+13}. 
The properties of this regime are reproduced automatically by our Boris-transport framework, as particles are propagated self-consistently, and only the largest magnetic bends, i.e., on  $\lc$-scales, should have an impact on particle propagation. 
While testing transport in this regime, we find good agreement with the trends shown in the Monte Carlo simulations from \citet{casse+01} and \citet{plotnikov+11}, i.e., the ratio of perpendicular to parallel diffusion falls as $(\rl/D)^{-2}$ as expected.

\subsection{Regime with $\rl < \lc$}\label{sec:propdiff}

For $\rl/\lc \ll 1$, the mean free path in turbulence with $\delta B/B \sim 1$ can also be expressed via Eq.~\ref{eq:lambda}, where $\delta$ is dependent on the phenomenology dictating the spectrum of 
turbulence \citep[e.g.,][]{bhattacharjee+10,treumann+15}. 
If scattering occurs via wave-particle resonance, the quasi-linear theory \citep[e.g.,][]{parker65,jokipii+66,forman+75,chapman+90}, returns $\delta =1/3 $ in the Kolmogorov regime \citep{kolmogorov41,stawarz+08} and $\delta =1/2 $ in the Iroshnikov-Kraichnan regime \citep{iroshnikov64}.

Recent advances in particle transport in turbulence have demonstrated that scattering on intermittent structures (magnetic field reversals at scales of the order of $\rl \leq \lc$) leads to a mean free path that can also be expressed based on Eq.~\ref{eq:lambda} \citep{lemoine23,kempski+23}, where $\delta \simeq 0.3$. 
In principle, $\delta$ may depend on the properties of intermittency \citep{lemoine23}, though a general theory has not been put forward, yet.

Since we only allow for magnetic structures at the coherence length scale, $\lc$, our calculations do not immediately capture the effects of resonant scattering or intermittency that could control the transport at smaller scales;
if left unaddressed, this limitation would lead to underestimate the transport delays for particles with $\rl \ll \lc$. 
To correct for this, we incorporate the influence of intermittency and resonant scattering on small scales by augmenting the propagation distance by a factor $(\lc/\rl)^{\delta}$, corresponding to the fact that for $\rl < \lc$ a particle traveling a distance of $\sim \lc$ generally undergoes $\lc/\lambda = (\lc/\rl)^{\delta}$ scattering. 
A posteriori, we find that varying $\delta \in [0.2, 0.6]$ has a negligible impact on our conclusions about the propagation delay of UHECRs.

\subsection{Particle Losses}

We consider the most important attenuation mechanisms, i.e., photomeson ($p\gamma$) and photodisintegration ($A\gamma$) interactions due to collisions with the cosmic microwave, infrared, and optical backgrounds (CMB, CIB, and COB, respectively).
We keep track of the evolution of the particles' atomic mass $A$ and charge $Z$ during propagation. 
For the heavier UHECRs, photodisintegration is the dominant loss process.

At every time step, an interaction probability is calculated for all species, along with a photodisintegration probability for heavier nuclei based on the Giant Dipole Resonance (GDR) cross section \citep[e.g.,][]{PANDORA22}. 
In a photodisintegration event, a photon-absorbing nucleus changes to another specie by releasing either a neutron or a proton, based on an interaction mean free path $\lambda_A$, as described, e.g., in \citet{mbarek+23}. 
Alpha particles can also be produced through photodisintegration, further depleting heavy nuclei of their nucleons. However, this channel is generally subdominant, and its contribution varies with the nuclear species involved. For example, the photodisintegration rate of oxygen into alpha particles accounts for only a few percent of its rate into protons \citep{gorbunov+63}. While this effect is element-dependent and could influence photodisintegration outcomes, its full impact lies beyond the scope of this work.

It is worth noting that the threshold energy for alpha production can, in some cases, be lower than for neutron or proton emission \citep{stecker+99}. However, the energy dependence of the corresponding cross section tends to be broad \citep[e.g.,][]{odsuren+15}, and modeling uncertainties—such as deviations in resonance—can further complicate predictions \citep{kido+23}.

A comprehensive treatment of particle propagation would ideally account for the effects of photo-nuclear interactions on particle rigidity, as these interactions modify the particle's energy and composition. Photodisintegration alters rigidity depending on the nucleon lost: losing a neutron reduces the rigidity from $\propto A/Z$ to $\propto(A-1)/Z$, while losing a proton increases it to $\propto(A-1)/(Z-1)$. 
For heavy nuclei (with $A\gg1)$ like iron, the loss of a single nucleon produces only a small change in rigidity and has minimal impact. In contrast, for lighter nuclei the effect is more significant. For instance, in nitrogen, losing a proton increases the rigidity by about 10\%, while losing a neutron decreases it by a similar amount. This results in a modest broadening of the rigidity distribution, centered around the initial value.

While our current results do not explicitly incorporate this effect, its influence can be estimated by summing the delays accumulated during each photodisintegration mean free path, treating the rigidity as constant between interactions and updating it after each event. Given that the overall change in rigidity remains small and symmetric around the initial value if interactions with protons and neutrons are equally probable, we do not expect this effect to substantially impact our conclusions.


\section{Results}\label{sec:general} 
Let us consider a source at fixed distance $D$ and different combinations of $\mathcal{R}$ and $\lc$, which induce a delay in particle propagation $t_d = t - t_b$, where $t$ is the total propagation time and $t_b$ is the ballistic travel time.
Figure~\ref{fig:gyro} shows the normalized time delay $t_d/t_b$, as a function of $\mathcal{R}$ and $\lc$ normalized to $D$.
Note that, since propagation only cares about $\mathcal{R} = E/(eZB)$, fixing the particle energy $E$ (or its rigidity, $\rho = E/Z$, where $Z$ is the charge) and choosing a value of the magnetic field are interchangeable. 
Particles are injected with an initial pitch angle $\mu$ to assess its effect on propagation.

In Figure~\ref{fig:gyro} we overplot the scaling of $\rl$ with $\lc$ using the definition of $\lambda$ from \S\ref{sec:rglc} and \S\ref{sec:propdiff}, where $\mathcal{R} \equiv \lambda^{1/\delta} \lc^{1-1/\delta}$ for the limiting case $\lambda = D$. 
Here, $\delta$ depends on the regime: $\delta \simeq 0.3$ for $\rl \ll \lc$, and $\delta = 2$ for $\rl \gg \lc$. 
The panels in Figure~\ref{fig:gyro} differ by the particle's initial pitch angle $\mu$, which is only relevant when $\lc > D$, where scattering is minimal but delay may still be non-negligible. 
In the following, we examine different regions of the maps with increasing delays (marked with letters a to d), delimited by gray lines in Figure~\ref{fig:gyro}:

\paragraph{\underline{Region a}}
In this region ($\mathcal{R} > D$) delays are minimal and associated with a slight bending from ballistic trajectories, such that $t_d/t_b \ll 1$.

\paragraph{\underline{Region b}} In this case $\lc > D$ and the B field remains coherent over the entire distance traveled (essentially a \emph{cosmic highway}), so no scattering is expected. 
As a result, for particles with $\mu = 1$, the propagation is nearly ballistic, and the associated delays are negligible. 
For $0 < \mu < 1$, though, a delay arises due to particle gyration and the total distance traveled can be approximated as $d \approx D \sqrt{\frac{\mu^2 + 1}{\mu^2}}$. 
Assuming an average pitch angle of $\sim \frac{\pi}{4}$, we expect delays of about $t_d/t_b \sim 0.7$, consistent with the numerical results shown in Figure~\ref{fig:gyro}.
For $\lc > D$, these calculations indicate that particle accumulate a delay of about 2 Myr for every Mpc separating us from the source. 
These delays alone may exceed typical AGN duty cycles, especially for sources on cosmological distance, though Mpc-long magnetic structures may be rare or even never realized.

\paragraph{\underline{Region c}} 
In this region $\lc < \rl < D$ and we see that $t_d/t_b$ is negligible above $\lambda = D$; instead, for sufficiently small values of $\rl/D$ and/or large values of $\lc/D \lesssim 0.1$, UHECRs are effectively scattered on scales much smaller than their Larmor radii, and hence potentially large delays may arise. 

\paragraph{\underline{Region d}} 
This is the part of the map where $ \rl < \lc < D$. Particles diffuse, and time delays are significant;
the initial pitch angle is irrelevant and time delays can easily be $t_d/t_b \gtrsim 10^3$. 
In this regime even the delay of UHECRs produced in circum-Galactic sources ($D \sim 30-300$kpc) would already exceed $>10^{2}$--$10^{3}$~Myr. 
Delays of several Gyr would likely photodisintegrate all of the heavy elements, at odds with Auger observations \citep[e.g.,][]{AUGER22,TA24}, as we discuss in \S\ref{sec:loss1}.

\begin{figure*}[htbp]
    \hspace{-2.1cm} 
\includegraphics[width=1.3\textwidth]{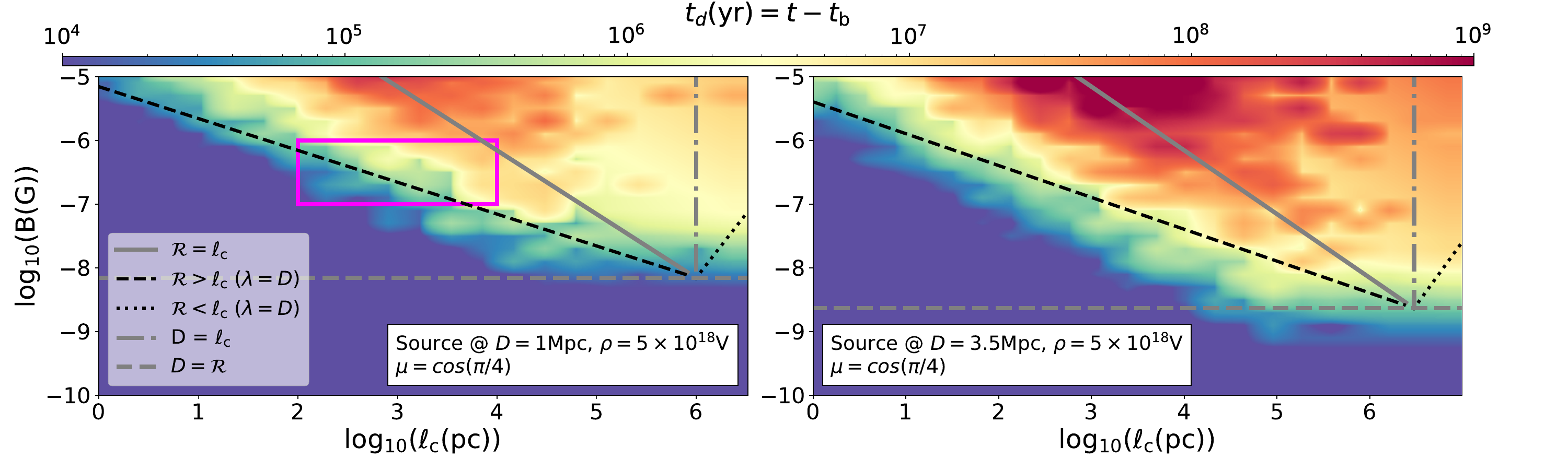}
    \caption{ 
        Time delay map $t_d$(yr) as a function of the coherence length $\lc$(pc) and the $B$-field $B$(G) that UHECRs with rigidity $\rho_{\rm max}=E/Z=5 \times 10^{18}$V probe over different distances as specified. Particles are initialized with $\mu = \cos{\pi/4}$, though the dependence on $\mu $ is minimal. The right panel corresponds to the distance to the nearest jetted AGN, Centaurus A.}
    \label{fig:time}
\end{figure*}

\section{Astrophysical Implications} 
The results presented above are general and apply to different parts of the journey of different CR species; 
we now discuss the implications for UHECRs in their sources and on galactic and extragalactic scales.

\subsection{General Considerations}

Figure~\ref{fig:time} shows the expected delay, $t_d$, for particle propagation as a function of coherence length $\lc$ and magnetic field strength $B$. 
Let us consider particles with rigidity $\rho_{\rm max} = 5 \times 10^{18}$ V over distances of 1 Mpc (the typical size of a galaxy cluster) and 3.5 Mpc, which corresponds to the distance to Centaurus A (Cen~A), a nearby AGN \citep[e.g.,][]{ferrarese+07,majaess10} with indications of an anisotropic UHECR excess above 40 EeV \citep{AUGER22}.
The rigidity $\rho_{\rm max}$ represents the highest rigidity needed to explain the UHECR flux and composition \citep[e.g.,][]{aloisio+14,auger17}, and thus minimizes the possible delay. 
For comparison, we also show the boundaries where $\lambda = D$ (introduced in \S\ref{sec:general} and Figure~\ref{fig:gyro}) as solid and dashed lines. 
When the delay becomes comparable with the age of the Universe, an effective \textit{magnetic horizon} arises \citep[e.g.,][]{parizot+04,globus+17b,globus+19,AUGER24b}, which has the effects of 1) limiting the observable Universe, 2) introducing a low-energy cut-off in the UHECR spectra, and 3) completely depleting the heavy elements.
These last two points may be important to account for the light chemical composition observed at $E\lesssim 10^{18}$eV and the relatively smooth transition between Galactic and extra-galactic CRs \cite[e.g.,][]{aloisio+12, unger+17}.

\subsection{Delays Due to Propagation in the Source\label{ssec:source}}
During the initial propagation phase, particles can spend a significant amount of time in regions inside or immediately around their sources.
In the case of AGNs, such regions (e.g., jet cocoons, lobes) can have sizes that reach 100kpc with $\mu$G fields.
From Figure~\ref{fig:gyro}, we can retrieve the maximum associated delays to be a few Myr. 
This heavily depends on the size of the regions, but could be another way in which particles are significantly delayed. 
For instance, tracking of accelerated particles in relativistic MHD simulations confirm that the cocoon is effective in isotropizing the UHECR distribution released by an AGN jet \citep{mbarek+19,mbarek+21a}.


\subsection{Propagation Inside Galaxy Clusters}
The left panel of Figure~\ref{fig:time} shows the delays expected on a distance of 1~Mpc, the typical size of a galaxy cluster, in particular for the typical cluster values of $B \lesssim 0.1-1 \rm \mu G$ and $\lc \sim 10^2-10^4 \rm pc$ (magenta box); 
such delays can already be comparable to or greater than the duty cycles of AGNs \citep[see also][]{berezinsky+97,blasi+98,matthews+21,condorelli+23}. 
This is particularly important since powerful jetted AGNs are predominately found in galaxy clusters \citep[e.g.,][]{begelman+84,best+07,fang+18}, and are prominent UHECR source candidates \citep[e.g.,][]{caprioli15,mbarek+19,katz+09,murase+19,jiang+21}. 
Also, over these timescales, a galaxy could change its identity and flip between AGN, starburst, or quiescent states, thus affecting any study that seeks a correlation with a specific class.

Propagation in galaxy clusters is most likely diffusive due to the presence of relatively strong magnetic fields, up to $\sim 1 \rm \mu G$, and coherence lengths on the scales of galaxy encounters (10-100kpc).
Within these environments, intermittent magnetic structures may play an important role, as they can significantly increase particle travel times, as described above \citep[also see][and references therein]{condorelli+23}. 
As a result, heavier nuclei are more likely to undergo frequent photodisintegration, leading to substantial changes in their chemical composition. 
We explore the effects of photodisintegration in more detail below.

\subsection{Centaurus A as a potential UHECR source}
In the right panel of Figure~\ref{fig:time}, we show the expected delays from particles emanating from Cen~A, which lies at a distance of $\sim3.5$ Mpc.
Provided that $\lc > 10^4$pc, which is not unrealistic on Mpc distances, UHECR delays could reach $\sim$1~Myr if the $B$-field $>$nG. 
On the other hand, if $B<$nG the delay would be comparable to $\sim$kyr due to the GMF \citep{alvarez-muniz+02, murase+09, takami+12}, always in addition to the delay inside the source (see \S\ref{ssec:source}).
Note that the delay calculated with a distance of 3.5 Mpc may be underestimated: it has been proposed scattering from the M81 or M82 group of galaxies \citep{bell+22} or due to the \textit{Council of Giants} around the Centaurus system \citep{taylor+23} may effectively produce a ``UHECR echo", which would imply a quite larger path length. 

While today Cen A looks like a rather weak FR-I, in the past it could have hosted a more powerful jet able to accelerate particles up to UHECR levels \citep[e.g.,][]{mbarek+21a,mbarek+25a}.
Delays $\gtrsim$ 1 Myr may then imply large uncertainties in the actual efficiency of Cen A at accelerating UHECRs (an argument valid for any source). For instance, a rather large fraction of Cen A's current bolometric luminosity would need to be converted into energetic particles, so that this AGN contributes a large fraction of the UHECR flux at Earth \citep{mbarek+25a}; this request would be eased if Cen A's jet were more luminous several Myr ago.

\subsection{UHECR Deflections During Transport}\label{sec:deflections}
Time delays go along with angular deflections in the apparent position of the source, an effect well investigated for the GMF \citep[e.g.,][]{takami+12b,jansson+12,CRpropa16,unger+17,unger+24,bister+24b}.
Even at the new of GMF deflections, one can only trace back UHECRs to sources if $\lambda \gg D$ and $\rl>\lc$. 

We can calculate trajectory deflection in our framework, and results are consistent with the calculations originally done in the diffusive regime, e.g., by \citet{waxman+96}.
If we assume a constant magnetic field magnitude, the average deflection will be $\langle \theta \rangle \sim \frac{\lc}{2\mathcal{R}}$ and the overall deflection $\theta_d$ will be $\theta_d \sim \langle \theta \rangle \sqrt{D/\lc}$ after a distance D.

In particular, we calculate that for a distance $D = 1$Mpc and $\rho = \rho_{\rm max}$, practically all particles with $\lambda \leq D$ are deflected by more than 20$^\circ$. 
As for particles with $\lambda > D$, 10\% are deflected by more than 10$^\circ$. 
This underscores the importance of including deflections due to extragalactic magnetic fields in propagation models when $\lambda \leq D$, an effect that adds on the time delay in hindering the correlation between present-day potential sources and UHECR arrival directions.

\begin{figure}
    \centering
    \includegraphics[angle=0,scale=.44,clip=true, trim= 0 0 0 0]{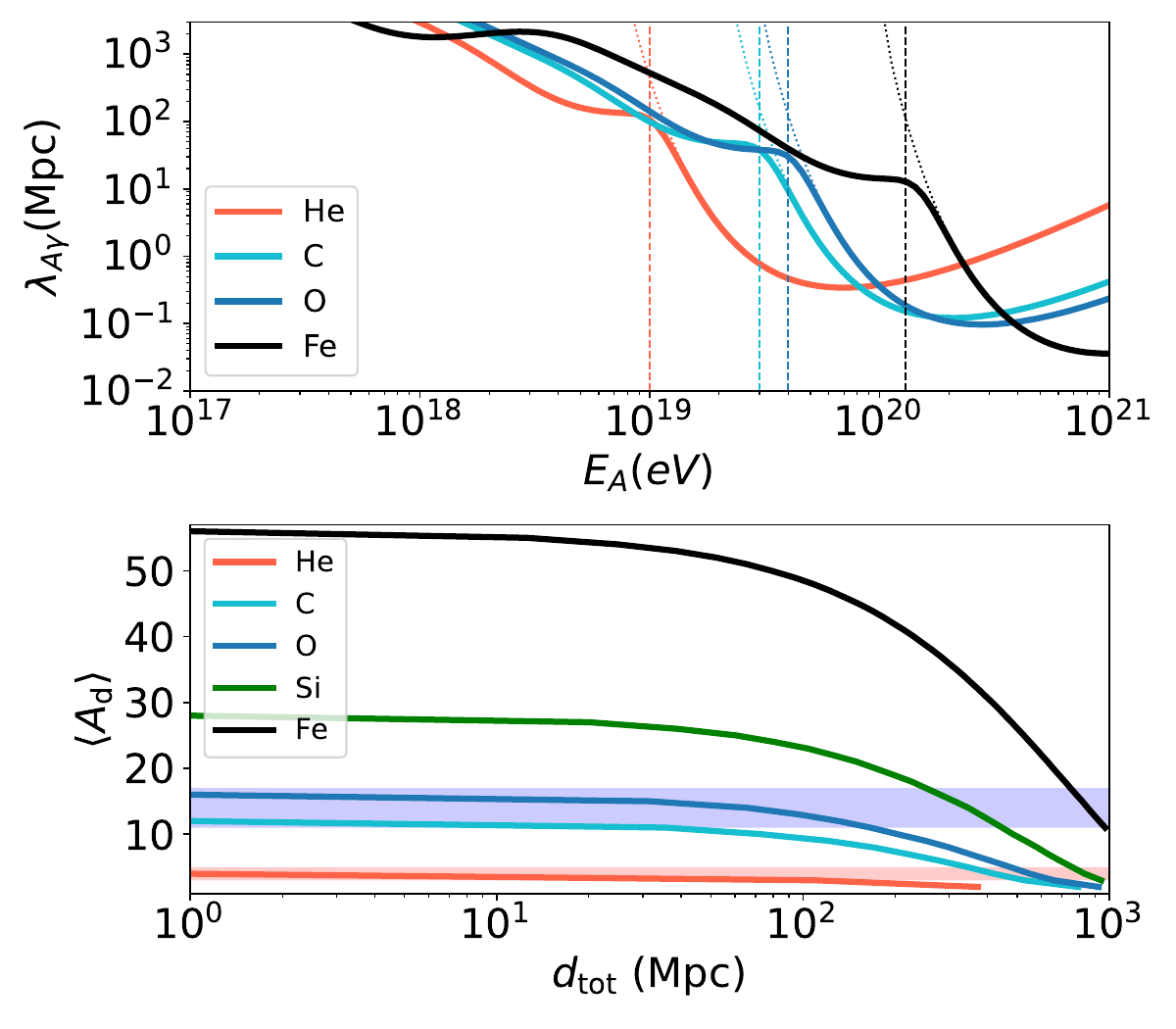}
    \caption{Upper Panel: Mean free path $\lambda_A$ for one photodisintegration interaction with CMB+COB+CIB due to the GDR for different species. 
    Dashed lines represent the maximum UHECR energy $5Z \times 10^{18}$eV. Dotted lines represent the contribution of the CMB only. Lower Panel: Maximum detectable atomic mass $\langle A_{\rm d} \rangle$ as a function of $d_{\rm tot}$, the total distance traveled for UHECRs with rigidity $\rho_{\rm max} = 5 \times 10^{18}$eV. The blue shaded area represents CNO-like detected particles ($\langle A_{\rm d} \rangle \in [11, 17]$), and the pink shaded area is for He-like particles ($\langle A_{\rm d} \rangle \in [3, 5]$).}
    \label{fig:lpg}
\end{figure}

\section{Effects of Losses}
Several studies have explored how photodisintegration processes influence the UHECR horizon \citep{aloisio+12, globus+23, AUGER24b}. In this work, we further these investigations by presenting some generalized constraints on the properties of the extragalactic magnetic fields, i.e., magnitude and coherence lengths. 
We also assess whether more precise measurements of heavy chemical compositions could refine current constraints.

\begin{figure}[tbp]
  \hspace{-.5cm}  
    \includegraphics[width=1.1\textwidth]{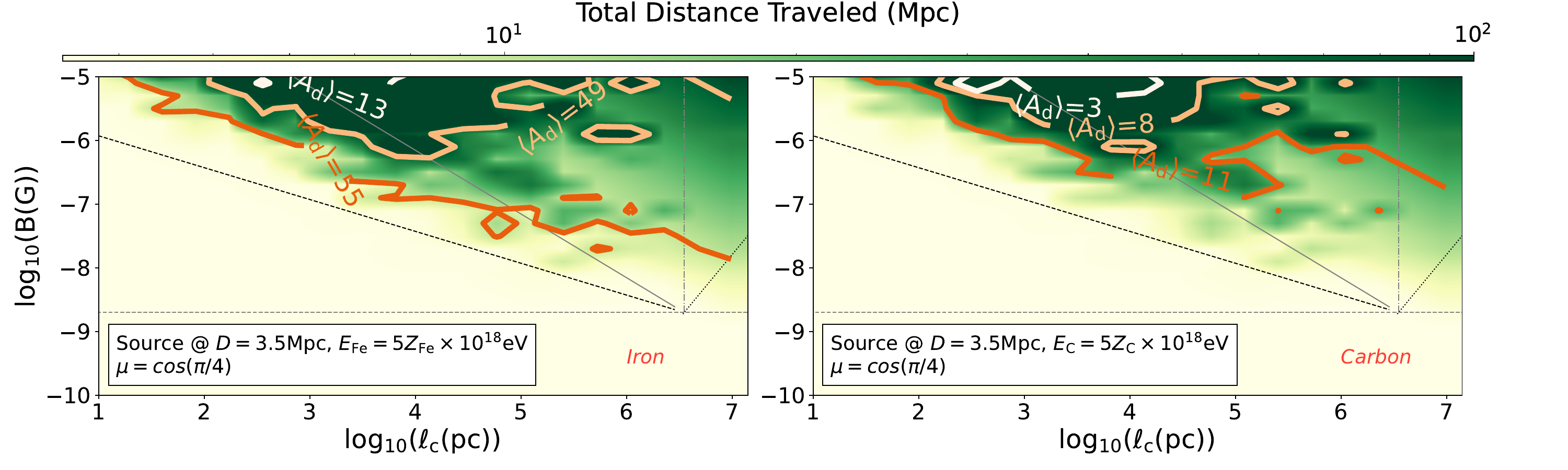}
    \caption{ 
        Map of the Distance traveled as a function of $\lc$(pc) and $B$(G) that iron (Fe) and carbon (C) UHECRs with rigidity $\rho_{\rm max}= E_{\rm Fe}/Z \simeq 5 \times 10^{18}$V probe. The contour plots show the maximum detected Atomic mass A of the parent UHECR (Fe or C) after propagation to Earth because of losses due to the CMB, CIB, and COB.}
    \label{fig:distance}
\end{figure}

\subsection{General Considerations}\label{sec:loss1}
The upper panel of Figure~\ref{fig:lpg} shows the mean free path $\lambda_A$ for different UHECR species with maximum rigidity $\rho_{\rm max} = 5 \times 10^{18}$eV. 
We find our GDR calculations to be consistent with previous studies including \citet{allard12}, {\tt SimProp} \citep[e.g.,][]{aloisio+12}, and {\tt CRPropa} \citep[e.g.,][]{batista+15,CRPROPA22}.
Ultimately, photodisintegration introduces a maximum detectable atomic mass $\langle A_{\rm d} \rangle$ for each injected heavy species from a particular source, which depends on the total distance traveled, $d_{\rm tot}$, as shown in the lower Panel of Figure~\ref{fig:lpg}. 

If UHECR sources are distributed across a range of distances, a fraction of the nuclei originating from nearby sources may arrive at Earth with their original atomic mass. 
As a result, this effect can extend the high-energy cutoff beyond what is expected from a purely distant source distribution, analogous to the energy suppression imposed by the Greisen--Zatsepin--Kuzmin effect \citep{greisen66, zatsepin+66}. 
This fraction of intact nuclei may contribute to a softer but detectable high-energy tail in the UHECR spectrum, reflecting the mixed composition and spatial distribution of the sources.

Before discussing our results, it is useful to lay out some preliminary considerations.
Since the reconstruction of the UHECR shower properties points to the most energetic UHECRs being as heavy as  CNO ($\langle A_d \rangle \sim 12$) \citep{Auger23}, we can devise two scenarios for the origin of such elements based on $d_{\rm tot}$. 
Either \emph{i)} detected CNOs are primaries that barely get photodisintegrated, which requires $d_{\rm tot} \leq$100~Mpc (see bottom panel of Figure~\ref{fig:lpg}), or \emph{ii)} they are secondary particles that come from photodisintegrated, Fe-like UHECRs, since Fe is likely the most abundant species among those with $A > A_{\rm CNO}$, as in Galactic CRs \citep[e.g.,][]{hoerandel+06,caprioli+10a, caprioli+17}. 

The scenario with secondary CNOs (from primary Fe-like) is quite fine tuned because it would imply \emph{i)} a dearth of He-like UHECRs, which are instead quite abundant, and \emph{ii)} that the total distance traveled by most, if not all, Fe-like nuclei must be exactly around $\sim$600-800~Mpc. 
An injection spectrum rich in Si-like nuclei could ease these requirements, but it is disfavored by standard nucleosynthesis channels.
The scenario with primary CNOs, on the other hand, seems more plausible, as the main requirement that needs to be satisfied is to have sufficient sources within $d_{\rm tot} \leq$100~Mpc. 

\subsection{Centaurus A as a potential UHECR source: Constraints From Losses}
In the following, we discuss a scenario where a significant fraction of the UHECR flux stems from Cen~A \citep[e.g.,][]{AUGER22}. 
Assuming that the UHECR injection composition includes CNO and Fe, Figure~\ref{fig:distance} shows a map of the traveled distance as a function of $B$ and $\lc$ from a source at $D = 3.5$ Mpc, the distance to Cen A. 
Overlaid on the map are the contours indicating the maximum detectable atomic mass, $\langle A_{\rm d} \rangle$, of Fe and C after propagation to Earth. 
For reference, we also include the limits defined in Figure~\ref{fig:time}.
We observe that a scenario where UHECR CNOs from Cen A are primary particles is feasible, with $A_{\rm d}(\text{C}) \simeq 11$--12 for $\rho_{\rm max}$ if $B \sim 0.1-1 \mu$G and $\lc \leq 10^4$ pc, as outlined above. 
Conversely, a scenario in which UHE CNOs from Cen A are secondaries (with Fe-like parent particles) would require the magnetic fields between us and Cen A to be unrealistically strong, $B \gtrsim 1 \mu$G (left panel of Figure~\ref{fig:distance}).
Furthermore, we note that the expected atomic mass of the parent Fe particles would be $A_{\rm d}(\text{Fe}) \simeq 55-56$, suggesting that detecting UHECRs with atomic masses $A \gtrsim 20$ would be plausible.

\section{Conclusions}
In this paper we considered the propagation of UHECRs across a wide range of magnetic field strengths and coherence lengths ($\lc$), examining the ensuing time delay, angular deflection, and attenuation losses as they travel toward Earth. 

$B$--$\lc$ maps (e.g., Figure~\ref{fig:gyro}) reveal that propagation delays can easily be comparable to, or even exceed, typical AGN duty cycles, particularly for coherence lengths $\lc > 100$ kpc.
Other important sources of delay, in addition to the $\sim$kyr expected from propagation in the Galaxy, are the scattering inside or around the sources, such as in the cocoon around an AGN jet \cite[see, e.g.,][]{mbarek+19, mbarek+21a}, and in galaxy clusters.
In particular, if UHECRs --- even at the highest rigidities, $\rho_{\rm max} = 5 \times 10^{18}$eV--- are produced in or pass through a Mpc-wide cluster with $B\sim 1\mu$G, delays can easily exceed the typical AGN duty cycles of $10^5-10^7$ yr (Figure \ref{fig:time}).

In general, large time delays and angular deflections may render it impossible to correlate UHECR arrival directions not just with short-lived sources (gamma-ray burst, tidal disruption events, newly-born ms pulsars), but even with sources with much longer duty cycles, such as AGNs and starbursts galaxies. 

Finally, we note how more precise measurements of the UHECR chemical composition at the highest energies, coupled with $B$--$\lc$ delay maps and photodisintegration calculations,  could constrain extragalactic magnetic fields, particularly when combined with observations of UHECR anisotropies that hint at potential sources such as AGNs or starburst galaxies \citep[e.g.,][]{Auger23}. 
For instance, the survival of Fe-like UHECRs from a known distance would put an upper limit on the average strength of extragalactic magnetic fields. 
Such a detection is a reasonable expectation, given the relative abundance of Fe in Galactic CRs and the observed trend of heavier UHECR composition at higher energies \citep{AUGER22}.

\section*{Acknowledgements}
We would like to thank Pasquale Blasi and Kohta Murase for their valuable feedback on the material covered by this work, and two anonymous referees for their comments, which helped improving the manuscript.
Simulations were performed on computational resources provided by the University of Chicago Research Computing Center.
This work was supported by NSF through grants AST-2009326 and AST-2308021.

\bibliography{MAIN}
\bibliographystyle{apsrev}

\end{document}